# Model Predictive Vehicle Yaw Stability Control via Integrated Active Front Wheel Steering and Individual Braking

Mümin Tolga EMİRLER, Bilin AKSUN GÜVENÇ

MEKAR Mechatronics Research Lab, Mechanical Engineering Department, İstanbul Technical University, Gümüşsuyu, Taksim TR 34437 İstanbul, Turkey

**Abstract**

Vehicle stability control systems are important components of active safety systems for road transport. The problem of vehicle lateral stability control is addressed in this paper using active front wheel steering and individual braking. Vehicle lateral stability control means keeping the vehicle yaw rate and the vehicle side slip angle in desired values. For this reason, a model-based controller is designed. The desired yaw rate is obtained from the single track vehicle model and the desired side slip angle is chosen as zero. Controller design consists of two parts, lower and upper controller parts. Upper controller is designed based on Model Predictive Control (MPC) method. This upper controller changes front wheel steering angles utilizing steer-by-wire system and also it generates the required control moment for stabilizing the yaw motion of the vehicle. Lower controller is an individual braking algorithm. It determines the individual braking wheel. In this way, the control moment can be applied to the vehicle. The designed controller is tested using the nonlinear single track vehicle model and the higher fidelity CarMaker vehicle model.

# 1. Introduction

Unexpected yaw disturbances caused by unsymmetrical vehicle perturbations like side wind forces, unilateral loss of tire pressure or braking on unilaterally icy road may result in dangerous lateral motions of a vehicle. Safe driving requires the driver to react extremely quickly in such dangerous situations. This is not possible as the driver who can be modeled as a high-gain control system with dead time overreacts, resulting in instability. Consequently, improvement of vehicle lateral dynamics by active vehicle control to avoid such catastrophic situations has been and is continuing to be a subject of active research [1, 2].

From literature, research papers may be categorized by three parts. These parts; only vehicle yaw rate control, only vehicle side slip angle control and integrated vehicle yaw rate and side slip angle control [3]. Besides, vehicle lateral dynamics control may be classified actuation type (steering or braking actuation) like active front steering control (AFS), active rear steering control (ARE), four wheel steering control (4WS), individual braking, active differential control.

Firstly, let's examine only vehicle yaw rate control papers. In [4-7], Ackermann et al. designed robust controllers for vehicle yaw rate control using only front wheel steering, only rear wheel steering or both of them. In these papers, the steering task of the driver separated two parts; path following and disturbance attenuation. These tasks separated using decoupling law and disturbance attenuation realized utilizing automatic control system.

Aksun Güvenç et al. used 2 DOF model regulator based control approach to solve vehicle yaw rate control problem and they found successful results in [2, 8]. They used active front wheel steering. Using this approach, controller was designed robustly to modeling errors and disturbance (unexpected yaw moment) rejection was realized. Designed controller was tested in hardware-in-the-loop simulations and experimental studies [9, 10]. Also, they tried to combine active front wheel steering and individual braking actuation together in their controller design [11-13].

Canale et al. designed vehicle yaw rate controller using internal model control approach, they used active rear differential actuation in their design [14, 15]. Moreover, they tried sliding mode control to solve vehicle yaw rate control problem [16]. Drakunov et al, also used sliding mode control and they realized their controllers using individual wheel braking [17].

Zheng et al used active front wheel steering and they developed a control algorithm following Ackermann's footsteps. They used a decoupling law similar to Ackermann's. They tested their design in experimental tests [18].

Only vehicle side slip angle control is generally realized with feedforward in literature [3].

Combined vehicle yaw rate and side slip angle control is the third and the last group of the literature research. Nagai et al, tried to control both (yaw rate and side slip angle) using active rear wheel steering and direct yaw moment control. They used state feedback control to control this MIMO system. They mentioned that state feedback gain could be found using control methods such as LQR, LQG and H infinity control [19, 20].

Yang et al, used integrated vehicle yaw rate and side slip control to solve the vehicle lateral control problem and they utilized active front steering and direct yaw moment control. When determining reference vehicle yaw rate and side slip angle, they used first order transfer functions. In this research, controller is designed as two parts: upper and lower controllers. Upper controller is a kind of optimal control named optimal guaranteed controller and lower controller is brake pressure distribution algorithm [21].

Falcone et al, propose a path following MPC based controller utilizing steering and braking. The control aim is to track a desired path for obstacle avoidance maneuver. The controller contains two parts: MPC based main controller and braking logic [22].

In this paper, we design an integrated controller different from literature using active steer-by-wire front wheel steering and individual wheel braking actuation. The controller separated two parts: MPC based upper controller and individual wheel braking based lower controller. The main contribution of this paper is (i) applying vehicle control using velocity scheduled MPC based control, (ii) realizing this control action utilizing steer-by-wire steering system and individual wheel braking, and (iii) testing designed controller in wide range of vehicle parameter (velocity, tire-road coefficient) and vehicle input (steering wheel angle) variations in high DOF simulation environment.

The organization of this paper is as follows. The vehicle models; single track vehicle model and CarMaker vehicle model, are introduced in Section II. The control strategy used in this paper is given in Section III. The controller design (MPC based upper controller and individual wheel braking based lower controller) is presented in Section IV. Section V is reserved for simulation studies, in this Section, are given along with nonlinear vehicle model and high fidelity CarMaker vehicle model simulation results are given. The paper ends with conclusions in Section VI.

## 2. Vehicle models

This section describes the vehicle models used for controller design and simulations. The two vehicle models used in this paper are the single track vehicle model and the higher fidelity CarMaker vehicle model.

The single track vehicle model is the simplest vehicle model that accurately captures lateral dynamics up to 0.3 – 0.4 g of lateral acceleration. In the single track vehicle model two tires on the same axle are lumped together and this results in one front an done rear tire set. Figure 1 shows the illustration and basic parameters of the single track vehicle model. The numerical values of vehicle parameters can be seen from Table 1. In simulations, these values are taken into consideration.

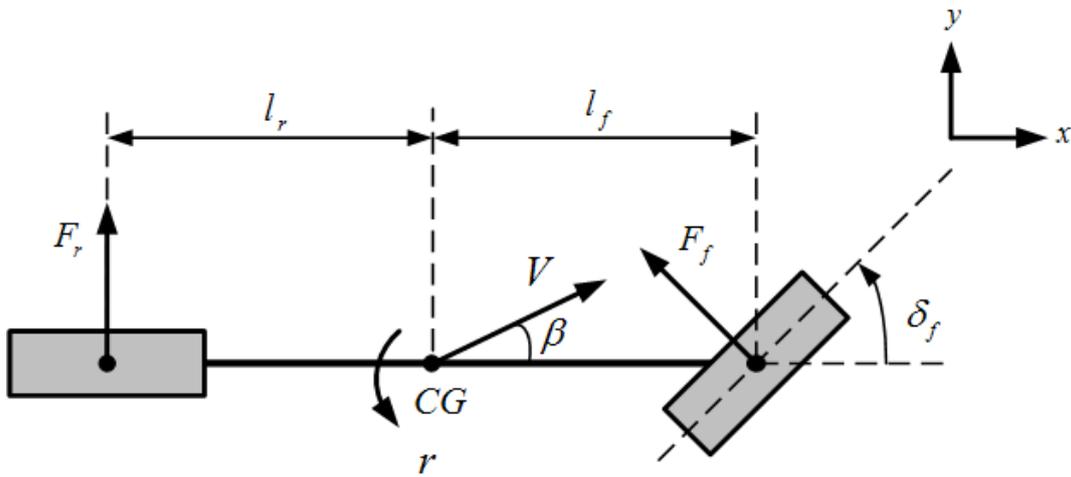

**Figure 1** Illustration of the single track vehicle model

**Table 1** Numerical Values of Vehicle Parameters

| | | |
|---|---|---|
| $m$ | 1321 | kg |
| $C_f$ | 72500 | N/rad |
| $C_r$ | 92500 | N/rad |
| $I_z$ | 2120 | kgm² |
| $l_f$ | 1.07 | m |
| $l_r$ | 1.53 | m |

The single track vehicle model contains nonlinear motion equations, which are the following:

$$F_x = m\left[\dot{V}\cos\beta - V\left(\dot{\beta}+r\right)\sin\beta\right] \quad (1)$$

$$F_y = m\left[\dot{V}\sin\beta - V\left(\dot{\beta}+r\right)\cos\beta\right] \tag{2}$$

$$M_z = I_z \dot{r} \tag{3}$$

In simulations, nonlinear single track vehicle model is used for testing of the designed controller. On the other hand, for the consideration of convenience and reducing the computational load, the model predictive controller design is worked out on the basis of the linearized single track vehicle model. Linearization is carried out some assumptions.

Assumption 1: The vehicle side slip angle is considered very small.

Assumption 2: The derivative of vehicle speed is taken as zero.

Assumption 3: The front wheel steering angles are considered very small.

Assumption 4: The wheel side slip angles are taken very small.

Under these assumptions, the linearized single track vehicle model is obtained. Its state space model is given as follows:

$$\begin{aligned}\dot{x} &= Ax + Bu \\ y &= Cx\end{aligned} \tag{4}$$

where

$$x = \begin{bmatrix}\beta & r\end{bmatrix}^T \quad u = \begin{bmatrix}\delta_f & M\end{bmatrix}^T$$

$$A = \begin{bmatrix}\dfrac{-(C_{ro}+C_{fo})\mu}{mV} & -1+\dfrac{(C_{ro}l_r - C_{fo}l_f)\mu}{mV^2} \\ \dfrac{(C_{ro}l_r - C_{fo}l_f)\mu}{I_z} & \dfrac{-(C_{ro}l_r^2 + C_{fo}l_f^2)\mu}{I_z V}\end{bmatrix},\ B = \begin{bmatrix}\dfrac{C_{fo}\mu}{mV} & 0 \\ \dfrac{C_{fo}l_f\mu}{I_z} & \dfrac{1}{I_z}\end{bmatrix},\ C = \begin{bmatrix}1 & 0 \\ 0 & 1\end{bmatrix}$$

Secondly, the designed controller is tested using CarMaker vehicle model. CarMaker vehicle model is a highly realistic model that incorporates simple engine dynamics, tire dynamics, steering dynamics, suspension dynamics, vehicle sprung body dynamics, longitudinal and lateral dynamics, a driver, road and environment models.

## 3. Control Strategy

The main control strategy is depicted Figure 2. The control strategy consists of upper controller (velocity scheduled model predictive control) and lower controller (individual wheel braking algorithm). This strategy employs active front wheel steering and individual brake torque distribution to control vehicle yaw rate and side slip angle.

The reference model is a 2 DOF linear single track vehicle model to produce the desired yaw rate. Besides, zero side slip angle is selected to obtain desired side slip angle. In reference model, tire-road friction coefficient (μ) is chosen as 1 to force the vehicle ideal driving behavior. Also, reference model was taken as velocity scheduled to eliminate the most important uncertainty source. The error between desired and measured yaw rate also the error between desired and measured side slip angle is feed to the velocity scheduled model predictive controller.

The velocity scheduled model predictive controller generates two output. One of them is front wheel steering angle ($\delta_f$) and the second one is corrective yaw moment (M). Considering that the front wheel steering angle is actuated by steer-by-wire steering system. The use of a steer-by-wire system, where the steering controller and driver steering commands are superimposed and sent to the steering actuator, is assumed in this paper. Moreover, corrective yaw moment transforms to the braking torques using lower controller and this actuation is realized by the braking system of the vehicle.

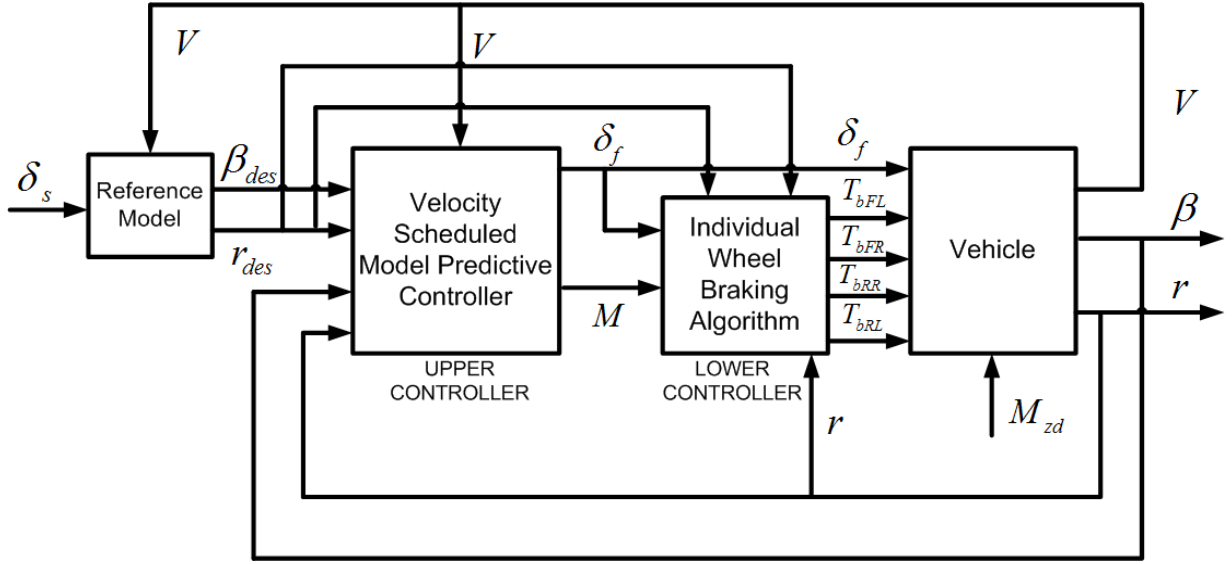

**Figure 2** Control strategy block diagram

## 4. Controller Design

### 4.1. Upper Controller – Model Predictive Controller

MPC based controller is designed to compute the optimal front wheel steering angle and corrective yaw moment in this section. The linearized vehicle model is used as prediction model of MPC, so that computational load of the calculations is reduced. The model in equation (4) is discretized by using ZOH method.

At each time k, the optimization problem below is solved,

$$J = \min\left[\sum_{i=1}^{p-1}(x_{k+i+1} - x_{k+i+1,ref})^T Q(x_{k+i+1} - x_{k+i+1,ref}) + \sum_{i=1}^{m-1}\left((\Delta u_{k+i})^T R_{\Delta u}(\Delta u_{k+i}) + (u_{k+i})^T R_u(u_{k+i})\right)\right] \quad (5)$$

subject to

$$x_{k+i+1} = A_D x_{k+i} + B_D u_{k+i} \quad (6)$$

$$\Delta u_{k+i+1} = u_{k+i+1} - u_{k+i}, \quad \forall i \leq m-1 \quad (7)$$

$$u_{\min} \leq u_{k+i+1} \leq u_{\max} \quad (8)$$

$$\Delta u_{min} \leq u_{k+i+1} - u_{k+i} \leq \Delta u_{max} \tag{9}$$

where $x_k = \begin{bmatrix} \beta \\ r \end{bmatrix}$, $x_{kref} = \begin{bmatrix} \beta_{ref} \\ r_{ref} \end{bmatrix}$, $u_k = \begin{bmatrix} \delta_f \\ M \end{bmatrix}$, $\Delta u_k = \begin{bmatrix} \Delta \delta_f \\ \Delta M \end{bmatrix}$

prediction horizon p = 15, control horizon m = 2, $T_s$ = 0.001 sec

$-15 \text{ deg} \leq \delta_f \leq 15 \text{ deg}$, $-1 \text{ deg} \leq \Delta \delta_f \leq 1 \text{ deg}$

$-10000 \text{ Nm} \leq M \leq 10000 \text{ Nm}$, $-100 \text{ Nm} \leq \Delta M \leq 100 \text{ Nm}$

$Q$, $R_u$ and $R_{\Delta u}$ are weighting matrices of appropriate dimensions.

The quadratic cost function and linear constraints yield convex optimization problem and this optimization problem can be solved using an Quadratic Programming solver (QP solver) such as Matlab Model Predictive Toolbox.

The constraints from system dynamics are mentioned by equations (4.2) and (4.3). The constraint equation (8) limits the front wheel steering angle and the corrective yaw moment. Moreover, the equation (9) limits the front wheel steering angle variation and the corrective yaw moment variation between two successive time steps.

The MPC based upper controller was designed considering velocity scheduled prediction model. Vehicle velocity was divided some intervals and for each interval new MPC based controller designed. These intervals can be seen from Table 2.

The if-else algorithm in Table 2 was programmed utilizing Matlab and it is used embedded with overall control design.

**Table 2** Controller Decision Algorithm

| Vehicle velocity [m/s] | Controller | Vehicle velocity used in prediction model [m/s] |
|---|---|---|
| $V < 25$ | Controller 1 | 20 |
| $25 \leq V < 35$ | Controller 2 | 30 |
| $35 \leq V < 45$ | Controller 3 | 40 |
| $45 \leq V < 55$ | Controller 4 | 50 |
| $45 \leq V < 55$ | Controller 5 | 60 |
| $V \geq 65$ | Controller 6 | 70 |

## 4.2. Lower Controller – Individual Wheel Braking Algorithm

Lower controller is a rule-based control algorithm based on individual wheel braking. After calculation of corrective yaw moment by upper controller, lower controller computes the individual wheel braking torques and determines which wheel will be braked.

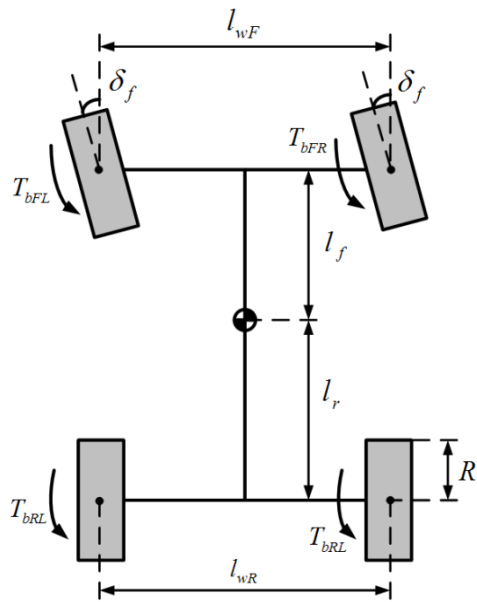

**Figure 3** Vehicle schematic

Braking torques for each wheels can be calculated using equation (10) and (11) for vehicle in Figure 3 [22].

$$T_{bFL} = T_{bFR} = \frac{|M|R}{\sin\left[\arctan\left((l_{wF}/2)/l_f\right) - \delta_f\right]\sqrt{l_f^2 + (l_{wF}/2)^2}} \quad (10)$$

$$T_{bRL} = T_{bRR} = \frac{|M|R}{\sin\left[\arctan\left((l_{wR}/2)/l_r\right)\right]\sqrt{l_r^2 + (l_{wR}/2)^2}} \quad (11)$$

The main differences of the equations for front and rear wheels arise from the existence of the front wheels steering angle.

While the algorithm development process, firstly all cases of vehicle yaw rate situation were derived. Then, in accordance with these cases, the braking wheel was decided. Six cases were found [21], these cases can be seen from Table 3. and also Figure 4.

Table 3  Vehicle cases for lower controller design

| Case | Vehicle yaw rate | Desired yaw rate | Situation | Braking wheel |
|---|---|---|---|---|
| 1 | $r > 0$ | $r_d \geq 0$ | $r_d < r$ | Front right (FR) |
| 2 | $r \geq 0$ | $r_d > 0$ | $r_d > r$ | Rear left (RL) |
| 3 | $r < 0$ | $r_d \geq 0$ | $r_d > r$ | Front left (FL) |
| 4 | $r > 0$ | $r_d < 0$ | $r_d < r$ | Front right (FR) |
| 5 | $r \leq 0$ | $r_d < 0$ | $r_d < r$ | Rear right (RR) |
| 6 | $r < 0$ | $r_d < 0$ | $r_d > r$ | Front left (FL) |

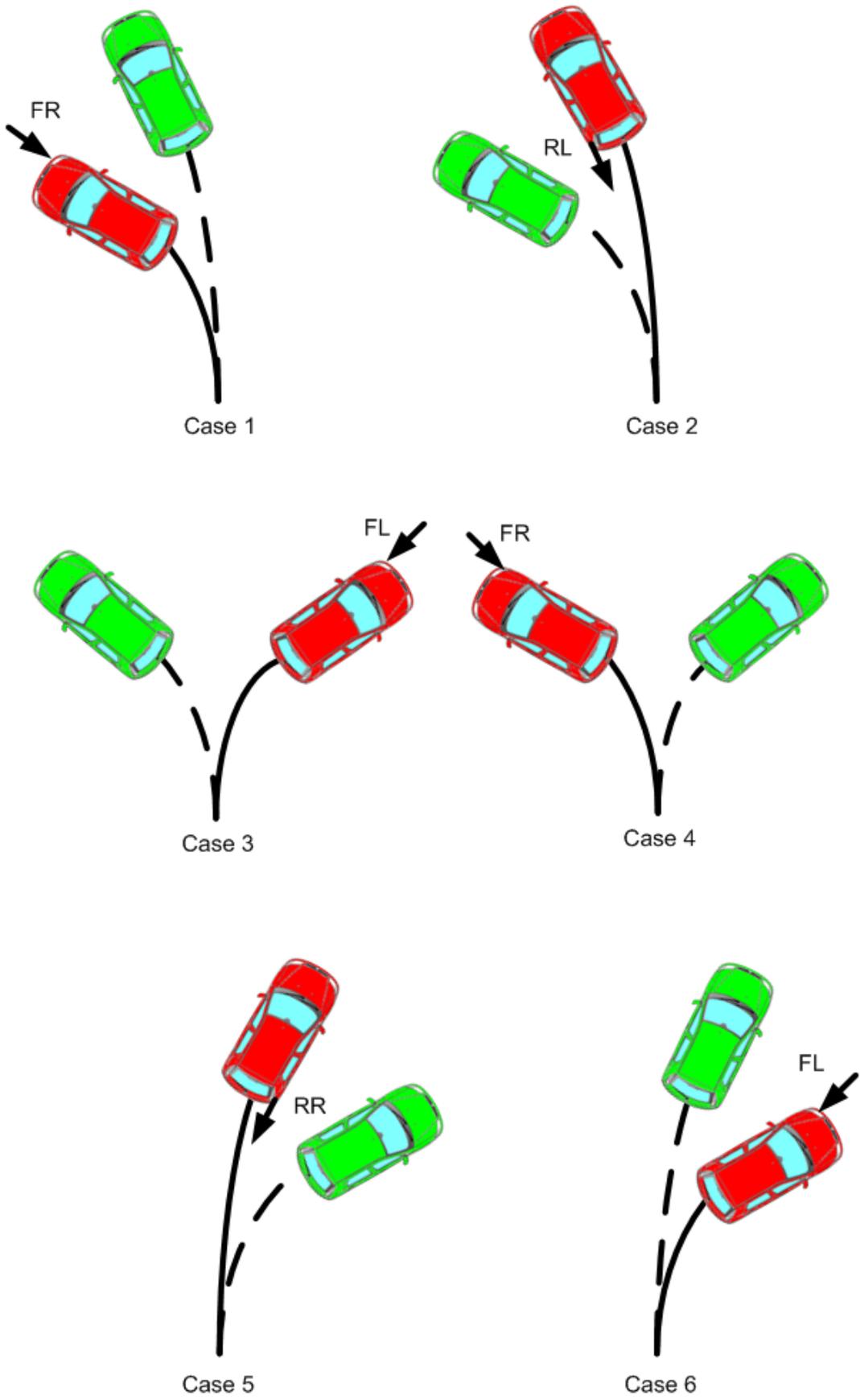

**Figure 4** Vehicle Cases

In Figure 4, red vehicle (normal path) shows the vehicle's first uncontrolled situation, and green vehicle (dashed path) shows the vehicle's situation after individual braking occurs. The abovementioned individual wheel braking algorithm was coded using Matlab and it runs embedded with model predictive controller in Simulink.

## 5. Simulation Studies

Designed integrated vehicle stability controller is tested using nonlinear single track vehicle model and the higher fidelity CarMaker vehicle model. In nonlinear vehicle model simulations, two different step steering wheel input and one step disturbance moment input were taken consideration separately. Simulation conditions were changed with manipulating vehicle initial speed and tire-road friction coefficient. So that, the designed controller was tested in a wide range. Besides, three stroscopic plots were drawn for different type of simulations to see vehicle trajectory changes. In CarMaker simulations, three different steering wheel input were used to obtain vehicle yaw rate and side slip angle response of the vehicle. Also, in these more realistic simulations, the controller was tested in a wide range of

## 5.1. Nonlinear STVM Simulation Results

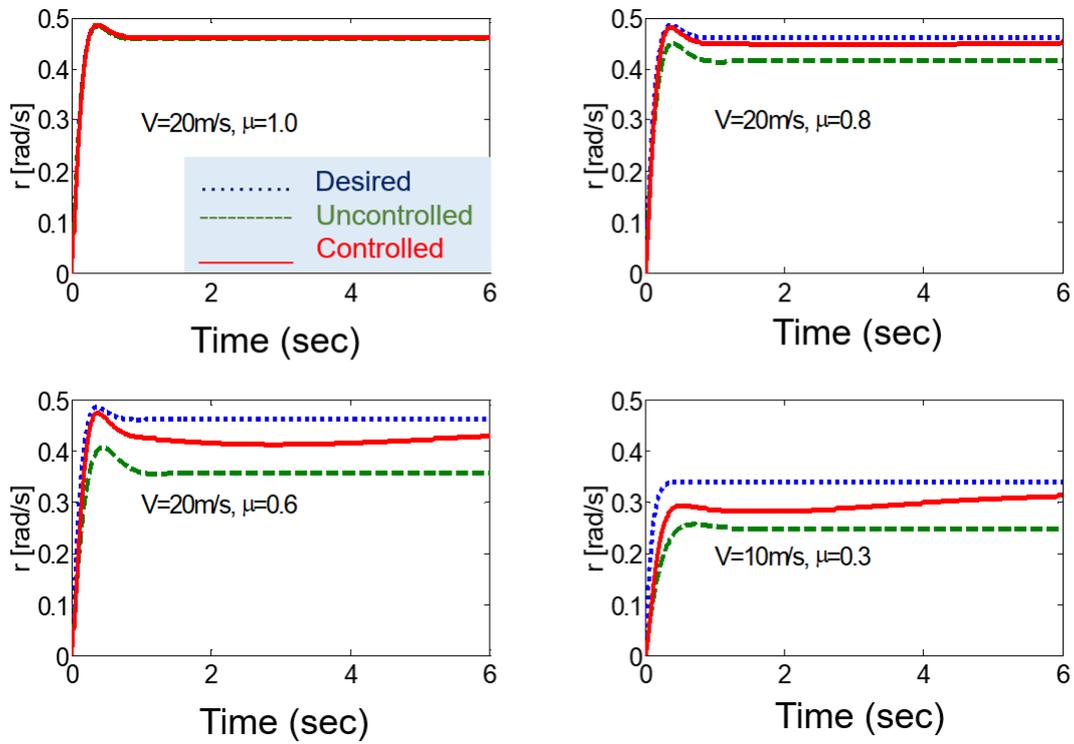

**Figure 5** Simulation 1, vehicle yaw rate response for $\delta_s = 90^o$, $M_{zd} = 0$.

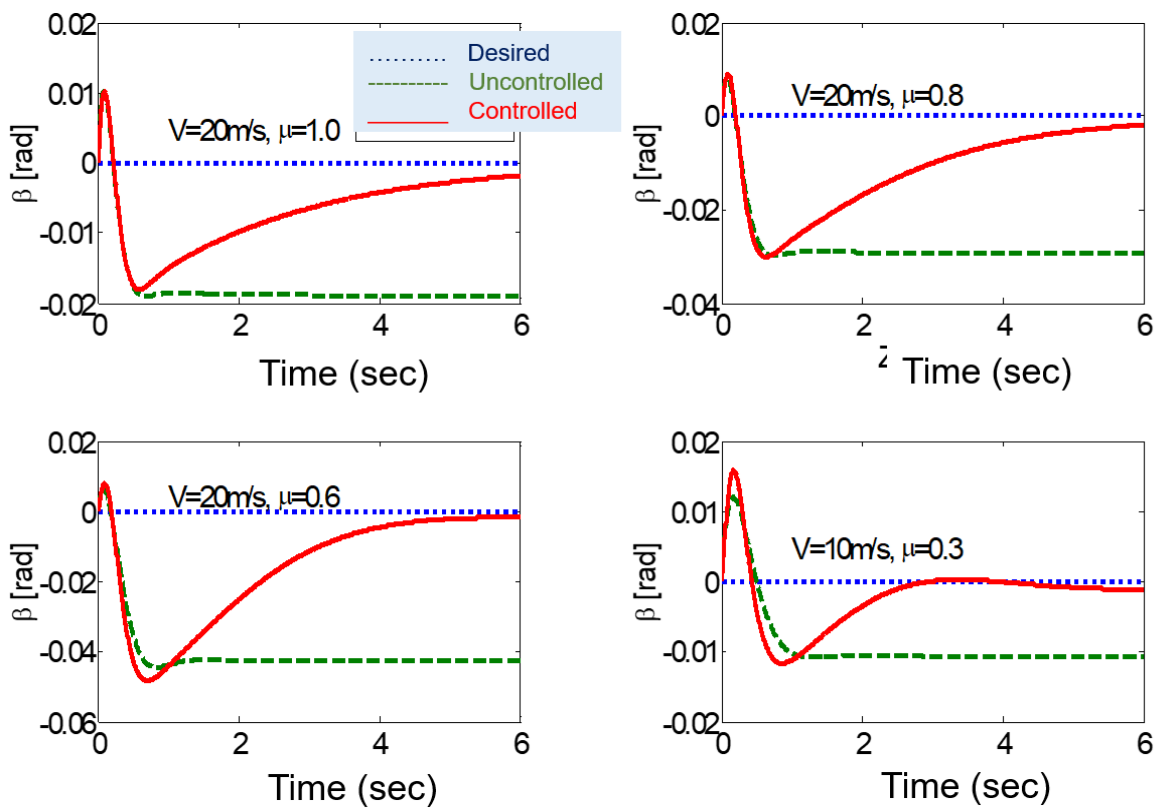

**Figure 6** Simulation 1, vehicle side slip angle response for $\delta_s = 90^o$, $M_{zd} = 0$.

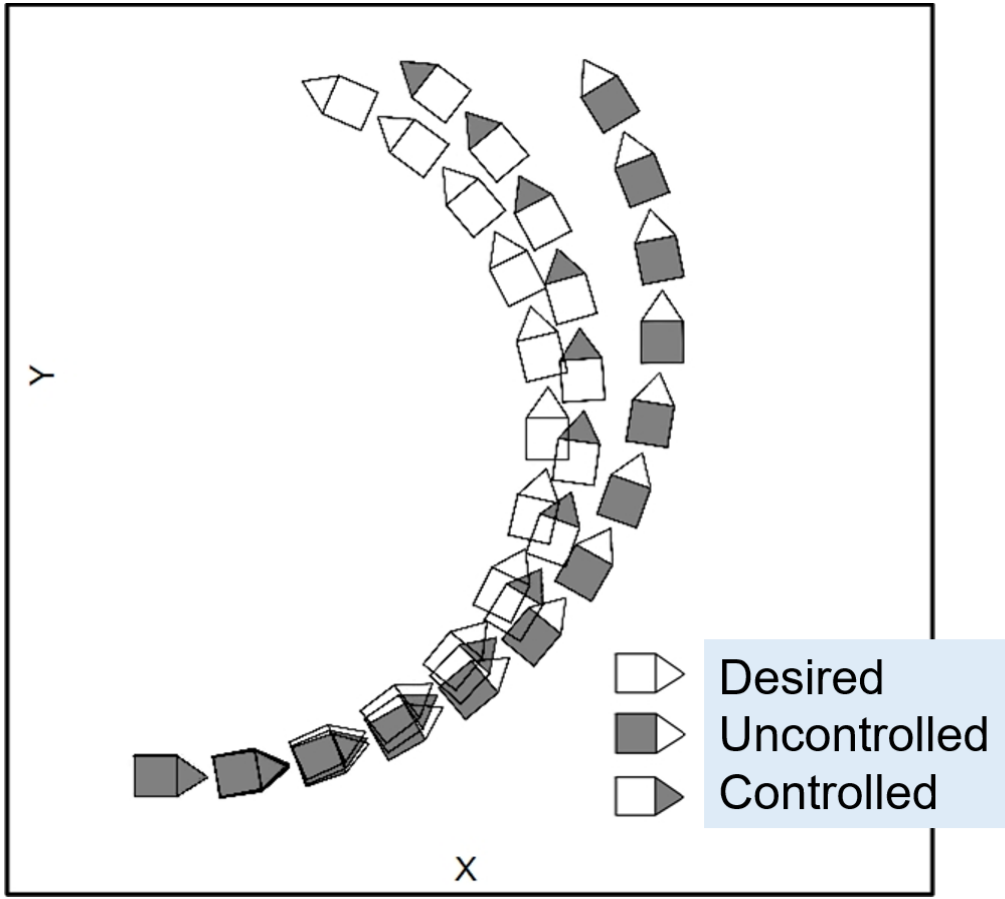

**Figure 7** Simulation 1, vehicle trajectory for $V = 20\ m/s$, $\mu = 0.6$

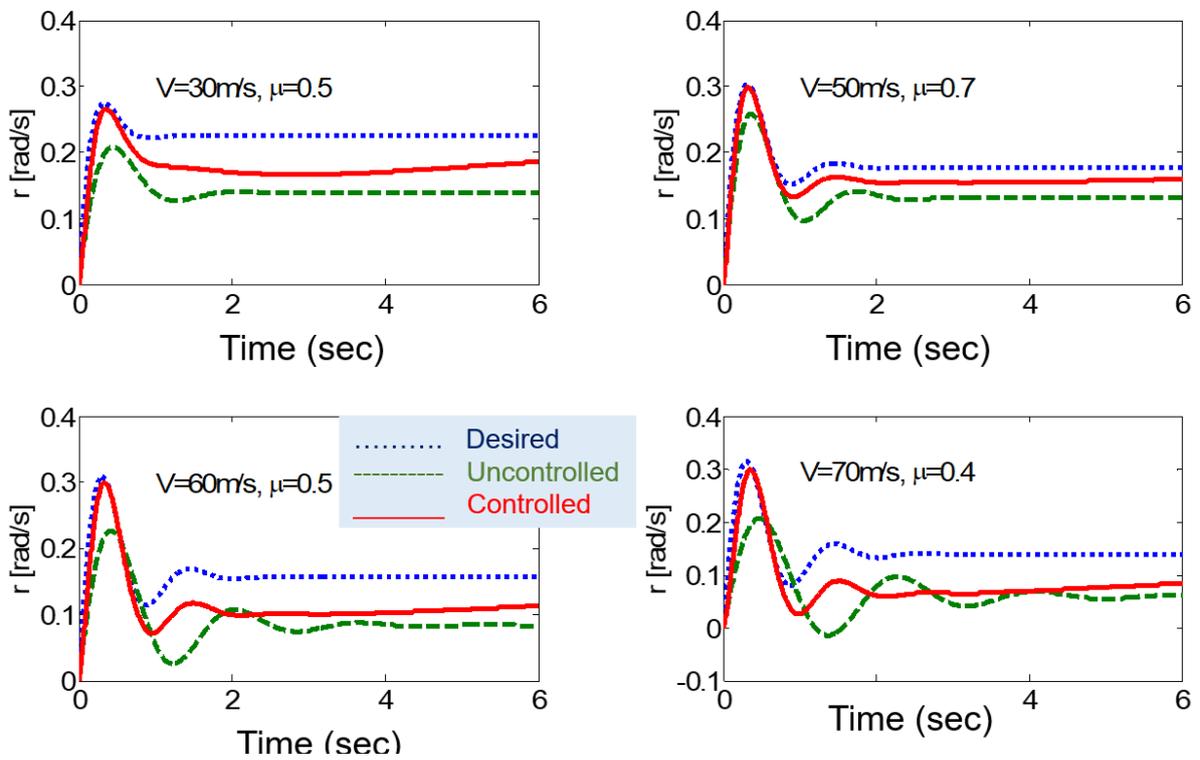

**Figure 8** Simulation 2, vehicle yaw rate response for $\delta_s = 45^o$, $M_{zd} = 0$.

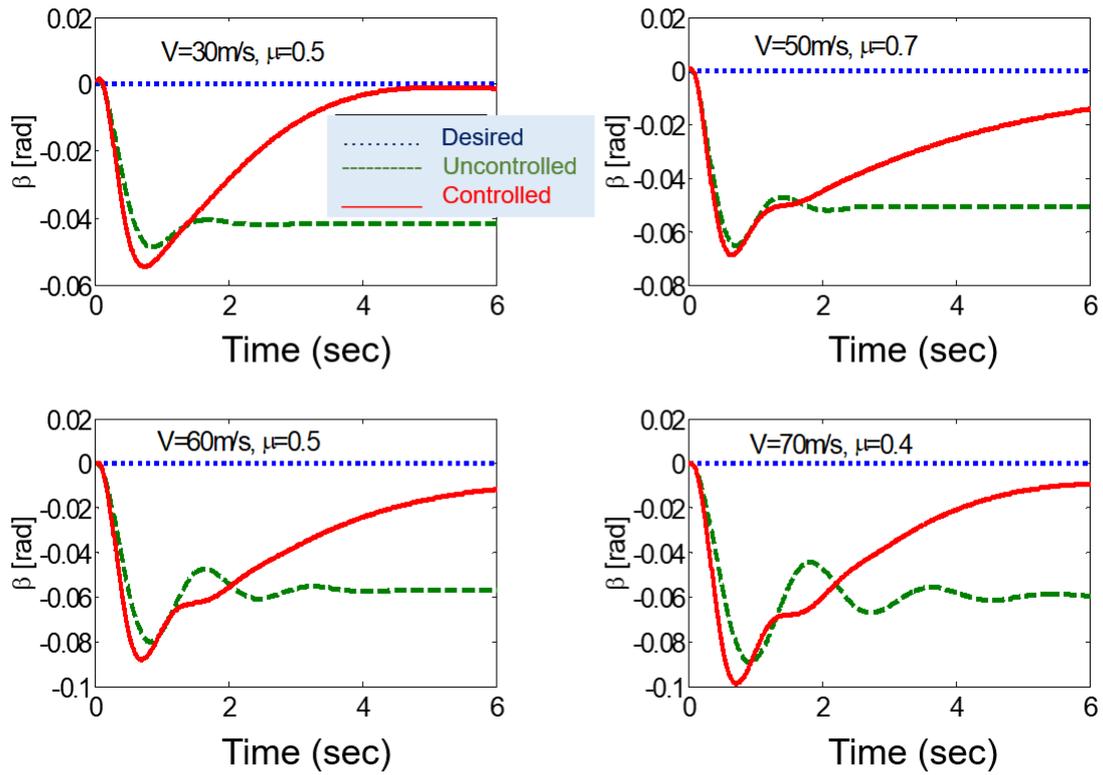

**Figure 9** Simulation 2, vehicle side slip angle for $\delta_s = 45^o$, $M_{zd} = 0$.

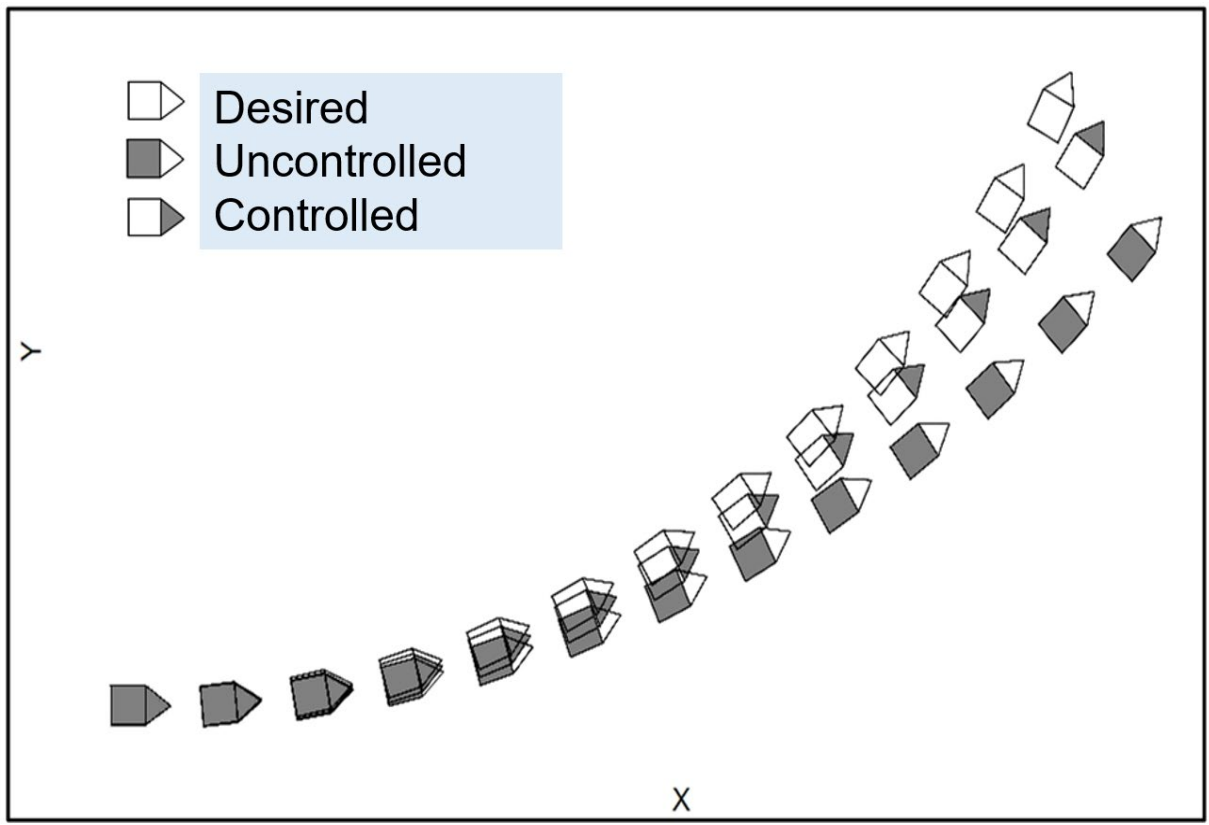

**Figure 10** Simulation 2, vehicle trajectory for $V = 50\,m/s$, $\mu = 0.7$

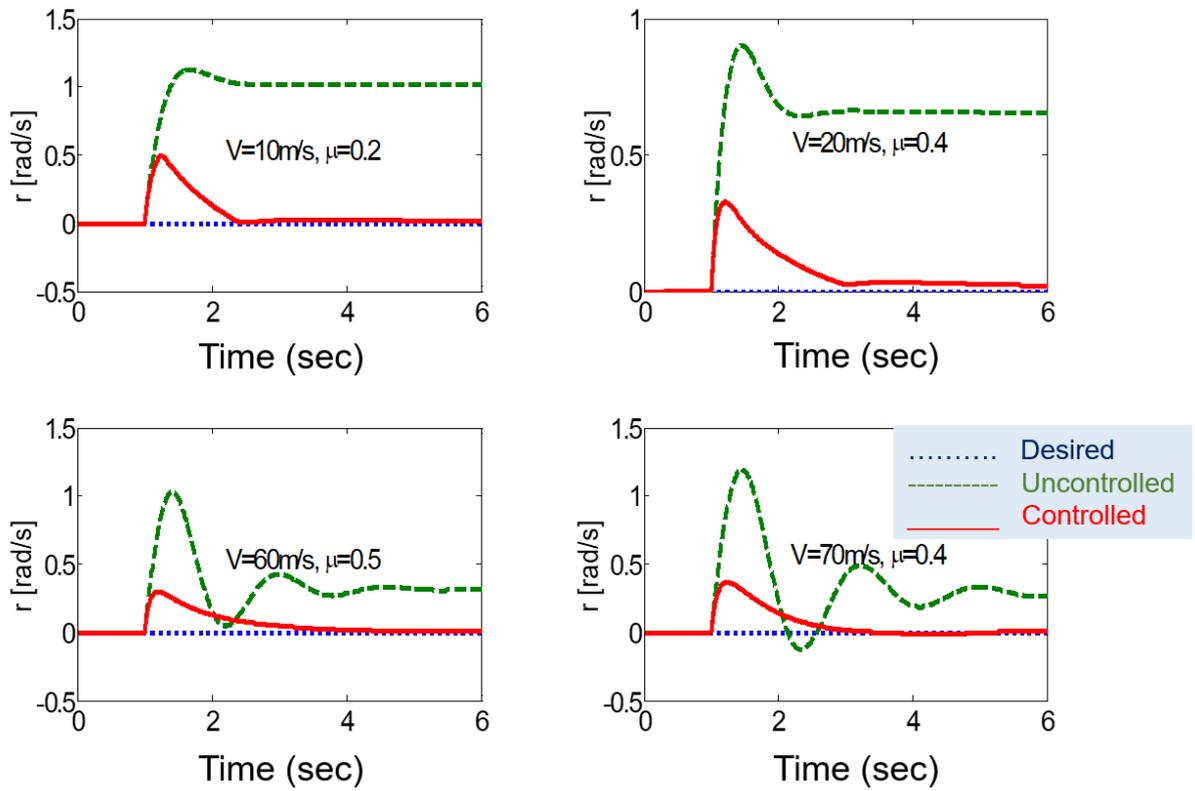

**Figure 11** Simulation 3, vehicle yaw rate for $\delta_s = 0$, $M_{zd} = 10000\ Nm$.

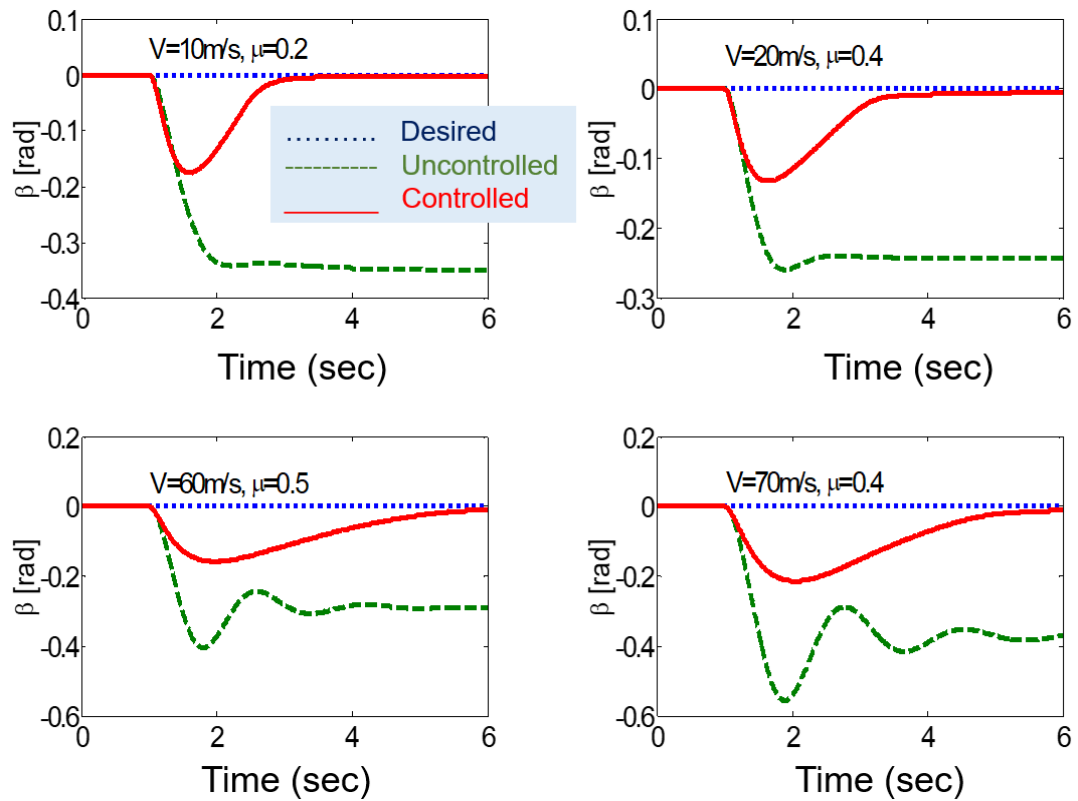

**Figure 12** Simulation 3, vehicle side slip angle for $\delta_s = 0$, $M_{zd} = 10000\ Nm$.

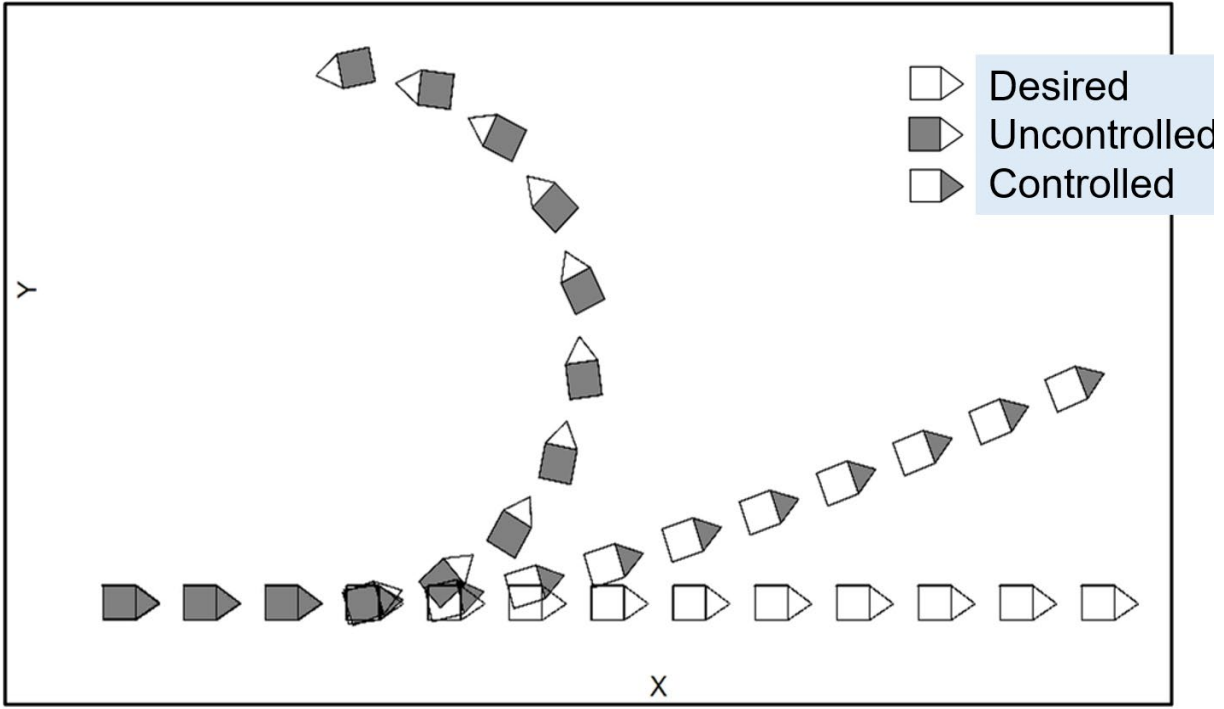

**Figure 13** Simulation 3, vehicle trajectory for $V = 70$, $\mu = 0.4$

## 5.2. CarMaker Vehicle Model Simulation Results

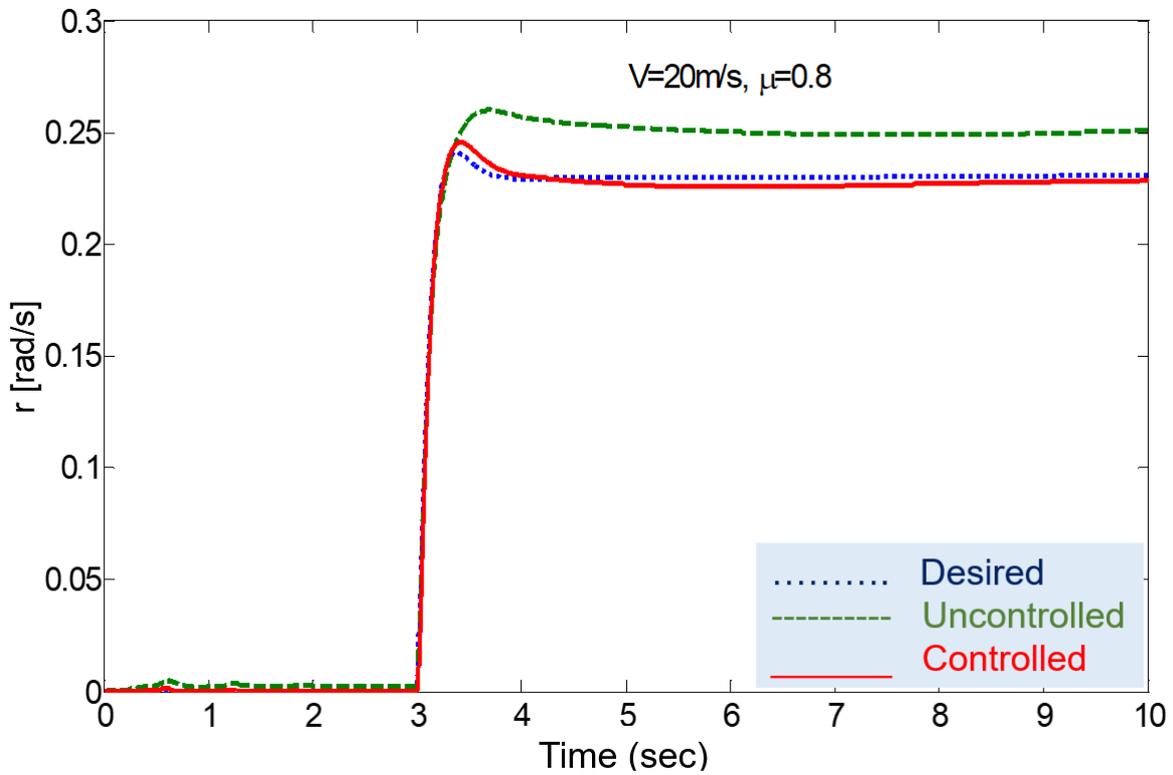

**Figure 14** CarMaker simulation 1, vehicle yaw rate response for $\delta_s = 45^o$, $M_{zd} = 0$.

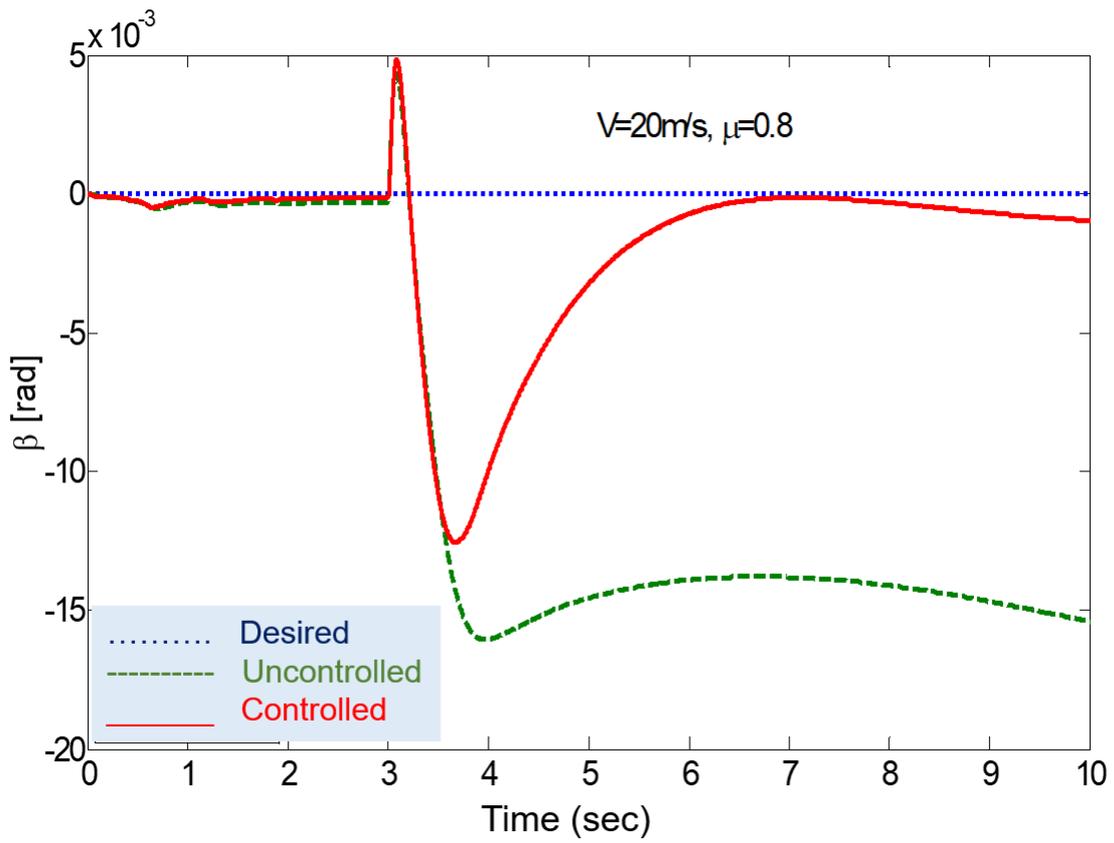

**Figure 15** CarMaker simulation 1, vehicle side slip angle response for $\delta_s = 45^o$, $M_{zd} = 0$.

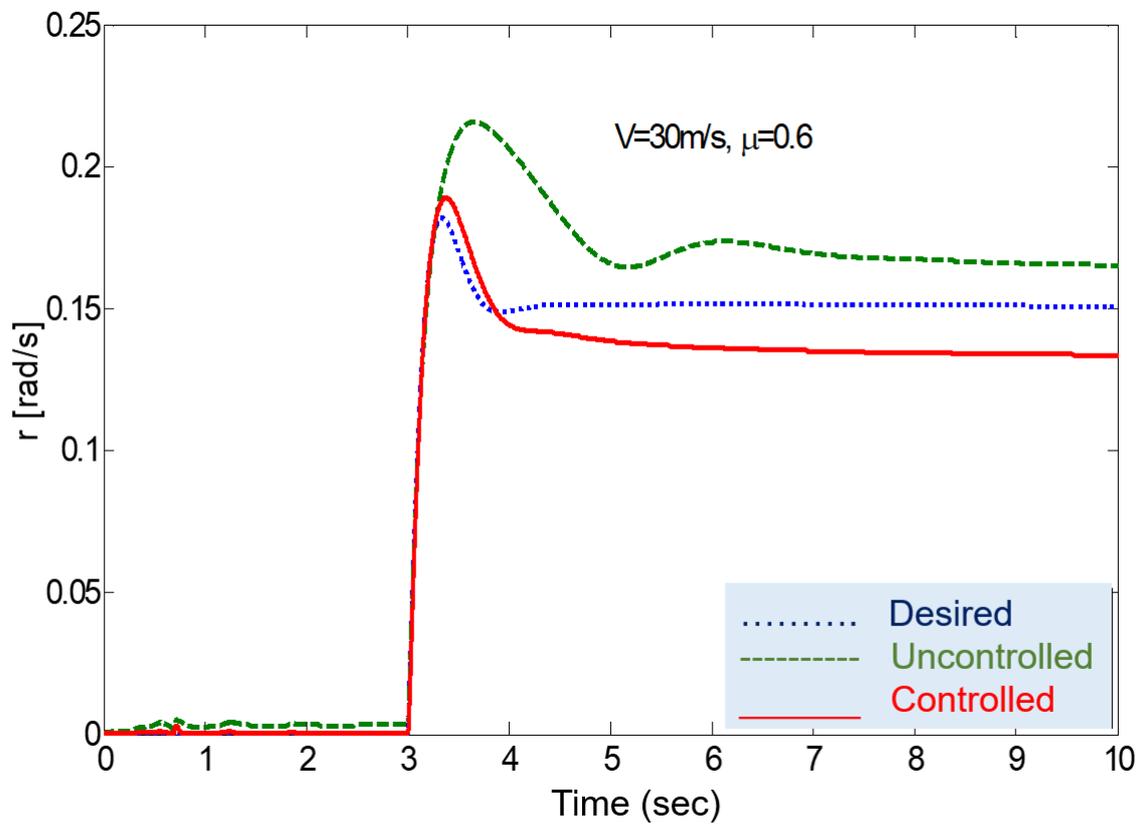

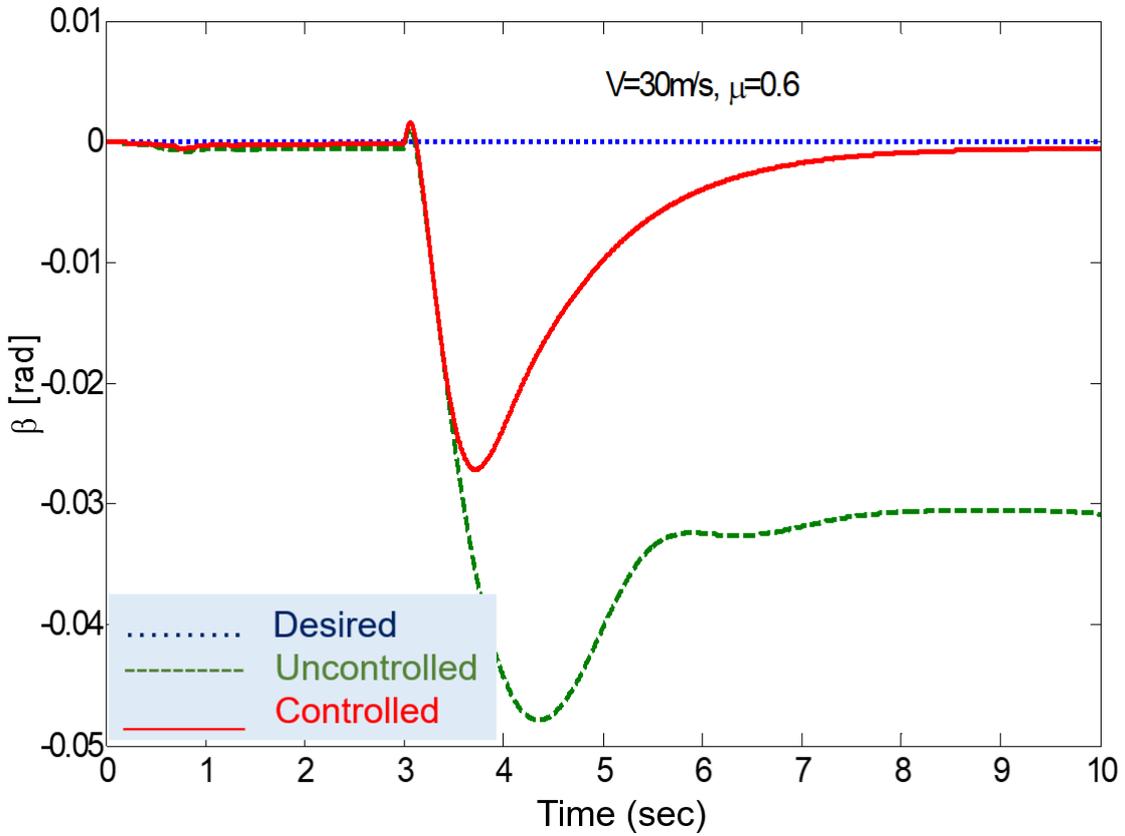

**Figure 16** CarMaker simulation 2, vehicle yaw rate response for $\delta_s = 30^o$, $M_{zd} = 0$.

**Figure 17** CarMaker simulation 2, vehicle side slip angle response for $\delta_s = 30^o$, $M_{zd} = 0$.

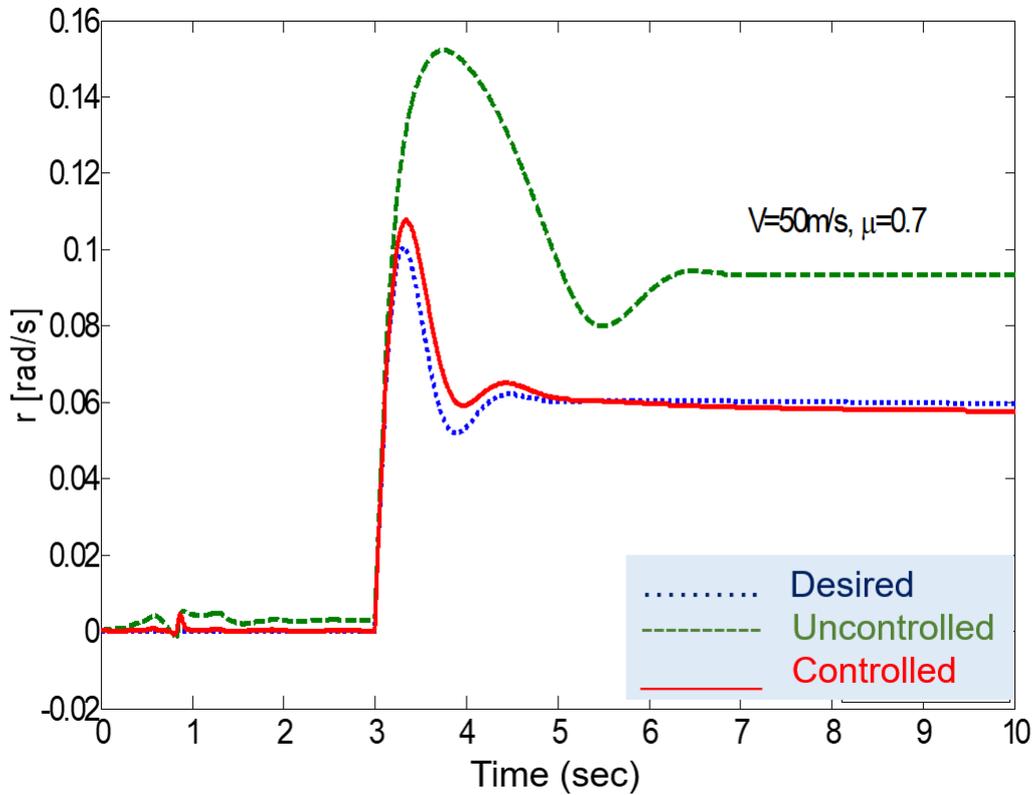

**Figure 18** CarMaker simulation 3, vehicle yaw rate response for $\delta_s = 15^o$, $M_{zd} = 0$.

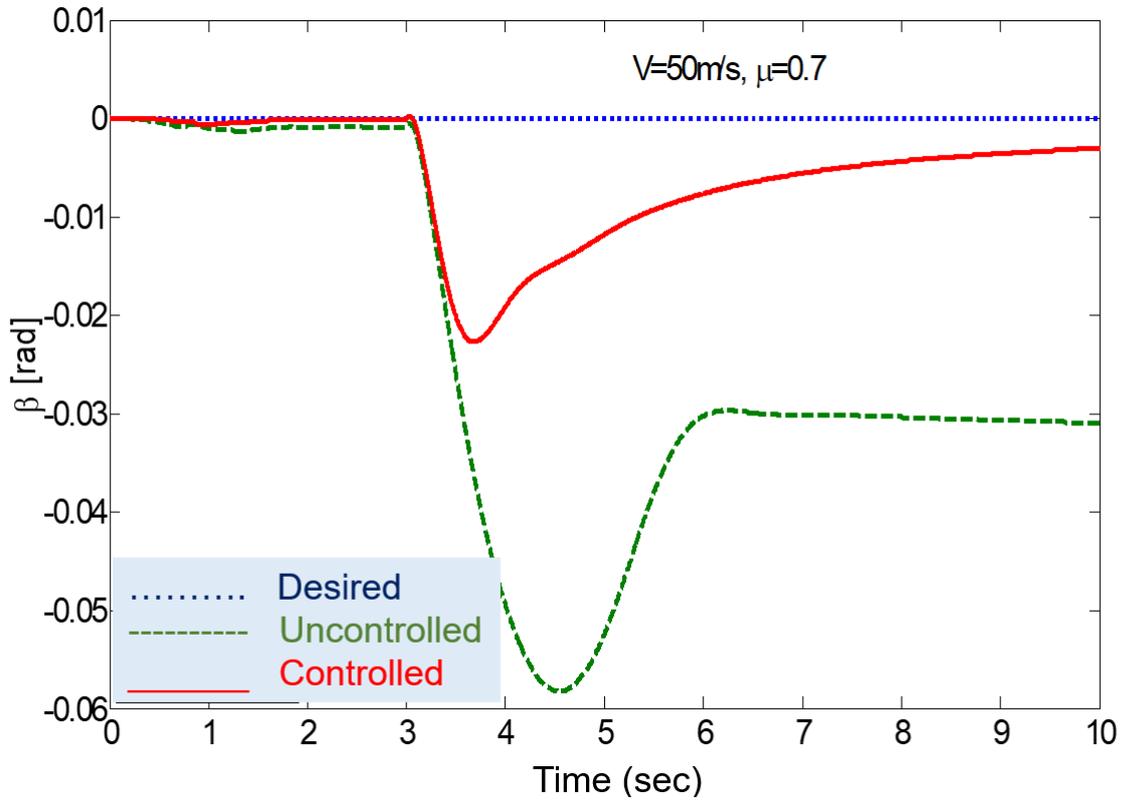

**Figure 19** CarMaker simulation 3, vehicle side slip angle response for $\delta_s = 15^o$, $M_{zd} = 0$.

## 6. Conclusions

In this paper, we proposed a solution to vehicle lateral stability control problem. We presented velocity scheduled MPC based upper controller and individual wheel braking based lower controller design. MPC based controller relies on a linear single track vehicle prediction model and the required constraints. Its optimization problem is solved by QP solver. The individual wheel braking algorithm distributes calculated corrective yaw moment to the wheels using braking torques for different cases and also it determines which wheels will be brake. Controllers realize actuation utilizing steer-by-wire steering system and braking system of the vehicle.

We tested our integrated (combined) controller with nonlinear single track vehicle model and the higher fidelity CarMaker vehicle model simulations. The simulations were performed on a wide range of different vehicle initial and tire-road coefficient alteration. The results showed that the use of integrated (combined) vehicle stability controller bring on encouraging results.

There have been many developments that were reported in the literature since the original writing of this paper. Other approaches like model regulation also called disturbance observer control (Oncu et al, 2007; Aksun-Guvenc and Guvenc, 2002, 2001; Guvenc and Srinivasan, 1995, 1994), speed scheduled LQR control (Emirler et al, 2015), intelligent control (Boyali and Guvenc, 2010), parameter space based robust control (Guvenc et al, 2017; Guvenc et al, 2021; Emirler et al, 2014, Emirler et al, 2015; Wang et al, 2018; Guvenc and Ackermann, 2001; Ma et al, 2021; Ma et al, 2020; Emirler et al, 2018; Zhu and Aksun-Guvenc, 2020; Zhu et al, 2019; Gelbal et al, 2020) and repetitive control (Demirel and Guvenc, 2010; Necipoglu et al 2011; Orun et al, 2009) for periodic speed profiles can also be applied for yaw stability control with application to fully electric vehicles. The yaw stability controller designed can also be tested as part of an autonomous driving system in an autonomous vehicle hardware-in-the-loop simulator (Gelbal et al, 2017; Cebi et al, 2005; Acar et al, 2019; Emirler et al, 2018; Hartavi et al, 2016; Emirler et al, 2016; Unal and Guvenc, 2014; ).

**Acknowledgements**

The first author would like to thank the support of TÜBİTAK National Scholarship Programme for PhD Students (TÜBİTAK BİDEB 2211).